\documentclass{aastex}          
\usepackage{spr-astr-addons}    
\usepackage{natbib}             

\begin{document}

\title{Impact of the short-term luminosity evolution on luminosity function of star-forming galaxies}

\shortauthors{S. L. Parnovsky}

\shorttitle{Impact of short-term luminosity evolution on LF of star-forming galaxies}

\author{S. L. Parnovsky}
\affil{Astronomical Observatory of Taras Shevchenko Kyiv National University\\
Observatorna str., 3, 04058, Kyiv, Ukraine\\
tel: +380444860021, fax: +380444862191\\ e-mail:par@observ.univ.kiev.ua}
\email{par@observ.univ.kiev.ua}

\begin{abstract}
An evolution of luminosity of galaxies in emission lines or 
wavelength ranges in which they are sensitive to the star formation process is caused by burning out 
of the most massive O-class stars during 
a few million years after a starburst. We study the impact of this effect on the luminosity function 
(LF) of a sample of star-forming galaxies.

We introduce several types of LFs: an initial LF after a starburst, current, 
time-averaged and sample ones. We find the relations between them in general and specify them in the case of 
the luminosity evolution law proposed for the luminous compact galaxies.
We obtain the sample LF 
for the cases the initial one is described by the pure Schechter function or the log-normal distribution and analyze
the properties of these LFs. As a result we get two new types of LFs to fit the LF of a sample of star-forming galaxies. 

\end{abstract}


\section{Introduction}\label{s:Introduction}

The luminosity function (LF) is a very important statistical characteristic of galaxy 
populations. The LFs provide information about galaxies and their evolution on the cosmological
timescale. In this paper we consider the LF of the galaxies with an active star formation 
in the wave range or emission 
line in which the lion share of radiation provides by a young stellar population. More
exactly we consider the radiation of galaxies shortly after a starburst. 
These galaxies underwent strong bursts of star formation on a 
time scale of a few Myr. Massive young stellar populations, which were formed
during the burst, emit plenty of ionising and non-ionising UV radiation.
Later on, the H$\alpha$ and UV luminosities strongly decrease with time 
after a starburst age $T_0\approx 3.2$ Myr, when the most massive 
stars start to fade out. 

Naturally, this is resulted in the changes of the observed LF. 
This fact affects the LF in
the wavelength ranges and emission lines which are good indicators of a star formation, while the LF in the optical range keeps its
form. This effect is mostly pronounced for the H$\alpha$, far ultraviolet (FUV), near ultraviolet (NUV) and far infrared (FIR) 
luminosities. We allow that taking into account the short-term luminosity evolution 
we can derive a more adequate form of the LF of a sample of 
star-forming galaxies.

The impact of the fading of galaxy luminosities on LF can be quite
significant. We describe it for the arbitrary initial luminosity distribution
and arbitrary luminosity evolution.
To study the impact of the factor under consideration we introduced the initial, 
the current, the time-averaged, and the sample LFs. In Section \ref{s:LF} 
we derive relations between all these LFs in general case of an arbitrary 
luminosity evolution. As a rule we consider galaxies with a single knot 
of star formation, but this LF can be applied to galaxies
with several knots of star formation, too. In section \ref{s:LF1} we specify these relations
for the luminosity evolution of luminous compact 
galaxies (LCG) discussed by \citet{PII}. 
In the sections \ref{s:aq} and \ref{s:ln} we use the popular forms of LFs, namely the Schechter function and the
log-normal one as the initial LFs and derive and analyze the corresponding 
LFs. Our findings are summarised in Section \ref{s:Sum}.

\section{Impact of the luminosity evolution}\label{s:LE}

The luminosity of the galaxy $L(T)$ is maximal one just after a starburst and
then decreases due to burning out of short-living massive stellar population:
\begin{equation}\label{e1}
L(T)=L_0f(T), \,\, f(0)=1,
\end{equation}
where $T$ is time after the starburst and $L_0$ is the luminosity at $T=0$.

The function $f(T)$ for the sample of 795 LCGs was derived by \citet{PII}.
This sample includes a very extreme population of compact galaxies with very high 
equivalent widths, representing the high-end tail of the LF of the overall star-forming population.
All galaxies in it underwent a burst of star formation less that 6 Myr ago.
So the radiation caused by the burst-like star formation prevails \citep[see discussion in][]{SBD}. 
\citet{PII} found that the luminosities in the H$\alpha$ emission line, FUV and NUV continuum are
described by the relation
\begin{equation}\label{e9}
f(T)=\left\{
\begin{array}{rl}
1&\textrm{if $T\le T_0$;}\\
\exp(-p(T-T_0))&\textrm{if $T > T_0$,}\\
\end{array}\right.
\end{equation}
with $T_0=3.2$ Myr and different values
of the parameter $p$ \citep[see details in][]{PII}. The same function is valid
for the luminosities in the 
22 $\mu$m IR band, calculated using {\it Wide-field Infrared Survey Explorer} 
(WISE) data and probably for the radio emission at 1.4 GHz \citep{AN}.

For generality we start with studying the case of arbitrary $f(T)$. 
We will specify the relation (\ref{e9}) later. We try to determine the effect 
of the short-term luminosity evolution on the observed LF of some sample of galaxies $n(L)$.
In order to do this we need some additional types of LFs. We introduce and consider
the initial $\eta(L_0)$, the current $\xi(L,T)$, the time-averaged $\varphi(L)$, 
and the sample $n(L)$ LFs. 

\subsection{The relations between initial, current, time-averaged and sample luminosity functions}\label{s:LF}

We assume that star formation in galaxies from a sample occurs in short
bursts separated by large periods without star formation,
which are much greater than the lifetimes of stars producing a main part of ionizing UV radiation.
Then, H$\alpha$, FUV and NUV luminosities greatly decrease before
the next consecutive starburst.
We also assume for simplicity that all galaxies have only single star-forming region.

The initial 
luminosity distribution $\eta(L_0)$ describes the distribution of the 
galaxy' luminosities $L_0$ immediately after the termination of the star 
formation burst. We consider all $L_0$ values for galaxies entering a unit volume
and describe their distribution by the function $\eta(L_0)$.
Here $\eta(L_0) dL_0$ is the number of starbursts per unit volume with the initial luminosity
in the interval from $L_0$ to $L_0+dL_0$.
The function $\eta(L_0)$ would coincide with the luminosity function 
$\varphi(L)$ if there were no temporal luminosity variations. 
This is a typical situation for the visible wavelength range,
in which the radiation from long-lived old stellar populations is dominated.
The similar situation occurs in the case of large galaxies with the continuous star formation,
when the individual bursts of star formation hardly affect the total luminosity.
On the other hand, the H$\alpha$, FUV and NUV luminosities in the LCGs are produced
by the radiation of the young stellar population \citep{PII}. Therefore they vary on short
time scales after the bursts of star formation.

After the time interval $T$ after the starburst the initial 
LF $\eta(L_0)$ turns into the time-dependent
current LF $\xi(L,T)$ because of decrease of galaxy luminosities.
Roughly speaking, the current LF is what the initial one becomes after time $T$.
The LF $\xi(L,T) dL$ is the number of galaxies per 
unit volume in the luminosity interval from $L$ to $L+dL$ at time 
$T$ after a starburst.

The galaxies from the sample are observed 
at different phases of luminosity evolution, 
i.e. with different time after the starburst $T$. \citet{PII}
have shown that the galaxies from the sample with $T<5.3$ Myr are
approximately uniformly distributed over starburst ages $T$. 
In this case an averaging over the sample 
corresponds to a time-averaging and therefore the time-averaged 
current LF $\xi(L,T)$ coincides with the LF $\varphi(L)$, which we
will call the time-averaged LF. Here $\varphi(L) dL$ is the number of galaxies 
per unit volume in the luminosity interval from $L$ to $L+dL$. 

Consider the case of monotonous luminosity decrease according to (\ref{e1}) with $df/dT<0$. 
Using the relationship (\ref{e1}) we find
\begin{equation}\label{e2}
\xi(L,T)=\eta(L_0)f(T)^{-1}=\eta\left( Lf(T)^{-1}\right) f(T)^{-1}.
\end{equation}
This current LF should be averaged over the time to find the 
function $\varphi(L)$:
\begin{equation}\label{e3}
\varphi(L)=\frac{1}{\Delta T} \int_0^{\Delta T} \eta\left( Lf(T)^{-1}\right) f(T)^{-1}dT.
\end{equation}
We apply the LF averaging over a time period $\Delta T$, which is much 
larger than the typical starburst age of the LCGs.
Instead of the time integration we apply an integration over $L_0$. 
In the case of constant $L$ we obtain the relation $f(T)dL_0=-L_0 \dot f(T) dT$
from Eq.~(\ref{e1}), where $\dot f(T)$ is the time derivative. Thus,
\begin{equation}\label{e4}
dT=-\frac{f(T) dL_0}{\dot f(T) L_0},
\end{equation}
where $dT$ is a time interval during which galaxies with initial luminosities 
from $L_0$ to $L_0+dL_0$ have the current luminosity $L$. Then
\begin{equation}\label{e5}
\varphi(L)=\frac{1}{\Delta T} \int_L ^{L_{\rm{max}}} \frac{\eta(L_0) dL_0}{L_0 |\dot f(T)|}.
\end{equation}
Here $L_{\rm{max}}$ is the maximum initial luminosity of galaxies in the sample. 
It can be assigned as $L_{\rm{max}}=\infty$ because 
$\eta( L_0)$ is a rapidly decreasing function at large $L_0$.

If a function $f(T)$ is constant at $T$ $\leq$ $T_0$ and  
decreases monotonously after $T_0$, Eq. (\ref{e5}) is transformed to
\begin{equation}\label{e6}
\varphi(L)=\left(\Delta T\right)^{-1} \left(\int_L ^{\infty} \frac{\eta(L_0) dL_0}{L_0 |\dot f(T)|}+T_0\eta(L)\right).
\end{equation}

The number of galaxies from the sample with the luminosity in the interval from $L$ to $L+dL$ is $Nn(L)dL$. Here $N$ 
is the number of galaxies entering the sample and $n(L)$ is a normalized LF of the sample.
To obtain the function $n(L)$ we multiply the function
$\varphi(L)$ by the volume $V(L)$ occupied by the sample galaxies with the 
luminosity $L$. The flux-limited sample includes galaxies up to the photometric
distance $r=(L/4\pi F)^{1/2}$, where $F$ is the minimal flux, if extinction is
not present. Considering flat space and neglecting the differences between 
the photometric distance and other types of distances in cosmology, 
we obtain that the volume $V(L)$ is proportional to
$r^3$ and correspondingly to $L^{3/2}$.

Thus, we assume $V(L)\propto L^{3/2}$ for flux-limited samples with small redshifts $z$. For the samples
with $z\gtrsim 1$ one ought to use the exact expression for $V(L)$ from the relativistic cosmology, e.g. from the $\Lambda$CDM 
model. Nevertheless, the simplified version with $V(L)\propto L^{3/2}$ has some advantage. Its simple form allows to 
obtain some useful relations like (\ref{eq2}).

For the flux-limited samples with small $z$ we obtain the sample LF
\begin{equation}\label{e7}
n(L)=\textrm{const}\times L^{3/2}\left(\eta(L)+T_0^{-1}\int_L^{\infty} \frac{\eta( L_0)
dL_0}{L_0 |\dot f(T)|}\right),
\end{equation}
where the constant is derived from the normalisation condition
\begin{equation}\label{e8}
\int_0^{\infty} n(L)dL=1.
\end{equation}

\subsection{Sample LF for the luminosity evolution law (\ref{e9})}\label{s:LF1}

Substituting $f(T)$ from Eq. (\ref{e9}) into Eq. (\ref{e7}) we obtain
\begin{equation}\label{e10}
n(L)=C\left(L^{3/2}\eta(L)+qL^{1/2}\int_L^{\infty} \eta(L_0) dL_0\right),
\end{equation}
where $C=\textrm{const}$ and
\begin{equation}\label{e11}
q=(T_0p)^{-1}.
\end{equation}
The parameter $q$ has a simple astronomical meaning. H$\alpha$ and UV
luminosities of a galaxy are constant during the period $T_0$, 
which is approximately equal to the lifetime of O stars. After that time
luminosities decrease by a factor of $\exp(-q^{-1})$ during every next time 
period $T_0$.

There are two marginal cases of Eq. (\ref{e10}). The luminosity function is 
reduced to the 
standard one ($q \to 0$) if $T_0 \to \infty$ (constant luminosities) 
or $p \to \infty$
(step, $f(T)=\theta(T_0-T)$, where $\theta$ is the Heaviside step function). 
The first term in
brackets in Eq. (\ref{e10}) vanishes ($q \to \infty$) if $T_0 \to 0$ 
(monotonous decrease without a plateau) or $p \to 0$ (very slow decrease). 
For the H$\alpha$ luminosity function
$p\approx 0.65$ Myr$^{-1}$ \citep{PII} and $q\approx 0.5$. 
For the FUV and the NUV luminosity functions the
values of $p$ are smaller and the values of $q$ are larger.  
Therefore, the initial and time-averaged H$\alpha$, FUV 
and NUV luminosity functions for star-forming galaxies are expected to be different.

Let us denote LFs for these cases as $n_1(L)=L^{3/2}\eta(L)$ and $n_2(L)=L^{1/2}\int_L^{\infty}x^{-3/2}n_1(x)dx$.
The function $n_1(L)$ is the LF of the sample without the luminosity evolution, i.e. in the case of constant
galaxy luminosities. Taking into account the luminosity evolution we obtain (\ref{e10}), which can be rewritten
in the form
\begin{equation}\label{eq1}
n(L)=C\left( n_1(L)+qn_2(L)\right).
\end{equation}
There is no point in considering the case $q<0$ as a real LF, but we can do this formally to analyze the properties 
of sample LF at $q\geq 0$. 
Let us denote the mean value of some quantity $F(L)$ over the sample as 
\begin{equation}\label{eq1a}
\langle F \rangle=\int_0^{\infty}F(L)n(L)dL.
\end{equation}
The relation
\begin{equation}\label{eq2}
\int_0^{\infty}L^sn_2(L)dL=\frac{1}{s+3/2}\int_0^{\infty}L^sn_1(L)dL
\end{equation}
is valid for $s=\textrm{const}$ if these integrals are convergent, in particular at $s\geq 0$. 
As a result we have the equation
\begin{equation}\label{eq3}
\langle L^s \rangle=\frac{1+\frac{2q}{3+2s}}{1+\frac{2q}{3}}\frac{\int_0^{\infty}L^sn_1(L)dL}{\int_0^{\infty}n_1(L)dL}
\end{equation}
connecting the sample averages of $L^s$ for LF $n$ taken into account affect of short-term luminosity evolution 
(\ref{e9}) and LF $n_1$ for the sample with invariable luminosities. Eq. (\ref{eq3}) at $s=1$ connects the 
sample average luminosities. At $s=2,3,\ldots$ it helps in calculation
variance, skewness, kurtosis, etc.

\section{LF for some standard distributions of initial luminosities after a starburst}\label{s:ss}

Let us try two popular functions often used as a functional form of LF as a function $n_1(L)$ in (\ref{eq1}).
The first one is the Schechter function  
\citep{ref:Schechter}, which is also known in the mathematical
statistics as the gamma distribution
\begin{equation}\label{sc}
\varphi(L)\propto (L/L^*)^{\alpha} \exp(-L/L^*).
\end{equation}
Parameters $\alpha$ and $L^*$ are determined from the shape of this function. 
The slope in the $(\ln \varphi(L), \ln L$) plane is equal to $\alpha$ 
for $L\ll L^*$ and changes rapidly at $L\approx L^*$.

According to \citet{G13}, a much better fit for LF of Galaxy And Mass Assembly (GAMA) 
galaxies with $z<0.35$ is provided by the Saunders distribution \citep{Sau}
\begin{equation}\label{saund}
\begin{array}{c}
n_1(L)=\textrm{const}\times u^{\beta}\exp\left(-A\lg^2(1+u)\right),\\
u=L/L^*_s,\quad A=1/(2\sigma^2).
\end{array}
\end{equation}
Here $\lg$ is the decimal logarithm and $\sigma, L^*_s=\textrm{const}$. Note that \citet{Sau} 
used the notation $L^*$ in (\ref{saund}). We add the subscript s to distinguish this quantity from the $L^*$ 
in (\ref{sc}). If the optimal value of $L^*_s$ in (\ref{saund}) is small and the majority 
of the sample galaxies have luminosities $L_i\gg L^*_s$, this distribution can be simplified.
Actually any value $L^*_s\ll \min (L_i)$ is acceptable. In this case we can drop the unity in the
$(1+u)$ term in Eq. (\ref{saund}) and we yield the 
two-parametric log-normal distribution
\begin{align}
\label{en1}
&n_1(L)=\left(\frac{a}{\pi}\right)^{1/2}\exp\left(-\frac{1}{4a}\right)\, \tilde{L}^{-1}\exp\left(-a\ln^2(L/\tilde{L})\right) \nonumber\\
&=\left(\frac{a}{\pi}\right)^{1/2}\exp\left(-\frac{1}{4a}\right)\,\tilde{L}^{-1}\left(\frac{L}{\tilde{L}}\right)^{-a\ln\left(L/\tilde{L}\right)}.
\end{align}
Here
\begin{equation}\label{en2}
a=A(\ln 10)^{-2},\quad \ln \tilde{L}=\ln L^*_s+\beta/(2a).
\end{equation}
In the marginal case $L_i\gg L^*_s$ the set of three parameters ($L^*_s,A,\beta$) can be redused to the 
the set of two parameters ($\tilde{L},a$). Only the combination of $L^*_s$ and $\beta$ gives the authentic 
parameter $\tilde{L}$. 

From (\ref{e7}) and (\ref{e8}) we can obtain the LFs for the sample of star-forming
galaxies with initial luminosities distributed according to the Schechter (\ref{sc}) and log-normal
(\ref{en1}) functions.

\subsection{Luminosity function for the Schechter distribution of initial luminosities}\label{s:aq}

Assume the initial LF is described by the pure Schechter function
\begin{equation}\label{e12}
\eta(L_0)\propto (L_0/L^*)^{\alpha} \exp(-L_0/L^*).
\end{equation}
We denote the ratio $u=L/L^*$ and substitute Eq. (\ref{e12}) into 
Eq. (\ref{e10}). The integral in
Eq. (\ref{e10}) becomes the incomplete gamma-function, \citep[see e.g. ][]{BE}
\begin{equation}\label{e13}
\Gamma(\alpha+1,u)=\int_u^\infty x^{\alpha} e^{-x} dx.
\end{equation}
The constant $C$ is obtained from the condition
Eq. (\ref{e8}). The final form of the distribution Eq. (\ref{e10}) with 
the condition Eq. (\ref{e12}) is 
\begin{equation}\label{e15}
n(L)=\frac{u^{\alpha+3/2}e^{-u}+qu^{1/2} \Gamma(\alpha+1,u)}{L^{*}(1+2q/3)\Gamma(\alpha+5/2)}, u=\frac{L}{L^*}.
\end{equation}
It has three parameters $q$, $\alpha$ and $L^*$. The relation (\ref{eq3}) 
is also useful for calculation
of the moments of this distribution, e.g. its mean value
\begin{equation}\label{e16}
\langle L \rangle =\frac{3}{5}L^*\left( \alpha+\frac{5}{2}\right) \frac{5+2q}{3+2q},
\end{equation}
variance, skewness, kurtosis, etc. Note that the mean value 
decreases with increasing $q$, as well as
\begin{equation}\label{e16a}
\langle L^s \rangle =\frac{3(3+2s+2q)}{(3+2s)(3+2q)}\frac{\Gamma(\alpha+s+5/2)}{\Gamma(\alpha+5/2)} (L^*)^s.
\end{equation}

\begin{figure}[tb]
\includegraphics[width=\columnwidth]{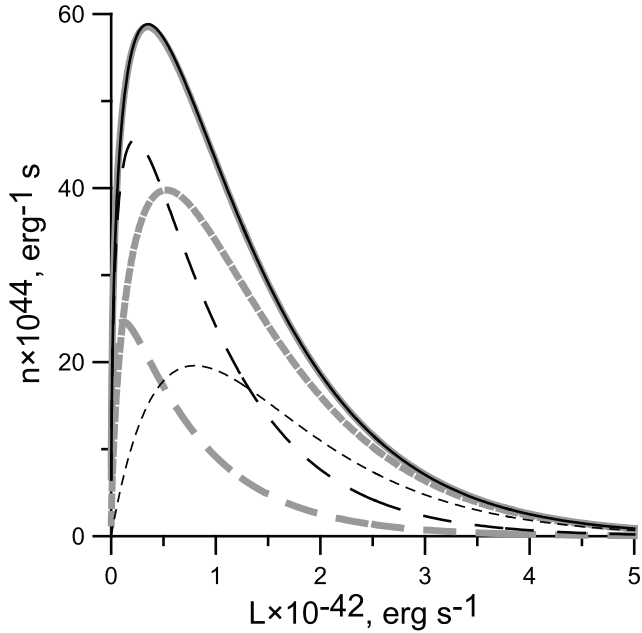}
\caption{The LF $n(L)$ (solid line) and the summands in (\ref{eq1}) 
(dashed lines with short and long dashes) for distribution (\ref{e15})
with two sets of parameters. 
Thick grey lines 
correspond to the set $q=0.48$, $L^*=8.5\cdot10^{41}$ erg s$^{-1}$, 
$\alpha=-0.88$ and thin black ones 
to the set $q=2.0$, $L^*=8.08\cdot10^{41}$ erg s$^{-1}$, $\alpha=-0.51$. 
Solid lines practically coincide. As a result, LF can be fitted by the functions (\ref{e15})
with two different sets of parameters. This fact complicates the approximation of LFs by this function}
\label{f1}
\end{figure}

The example of this LF one can see in Fig. \ref{f1}. We plot the graphs of dependence of $n(L)$ (solid line),
$n_1$ (short-dashed line) and $n_2$ (long-dashed line) vs $L$ for two sets of parameters. Thick grey lines 
correspond to the set $q=0.48$, $L^*=8.5\times10^{41}$ erg s$^{-1}$, $\alpha=-0.88$ and thin black ones 
to the set $q=2.0$, $L^*=8.08\times10^{41}$ erg s$^{-1}$, $\alpha=-0.51$. These sets are obtained by fitting
the LF of the sample of about 800 LCG, used by \citep{PII}.

We see that these two different sets provide practically the same LFs $n(L)$. Naturally,
this fact causes some problems for fitting the LF obtain from observations by function 
(\ref{e15}). Let us show that this situation is typical. Suppose we know the values of
$\langle L \rangle$, $\langle L^2 \rangle$ and $\langle L^3 \rangle$ for the function (\ref{e15}).
We can obtain the set of parameters $q$, $\alpha$ and $L^*$ for this case.
From (\ref{e16}) we have
\begin{equation}\label{eq4}
L^*=\left( \alpha+\frac{5}{2}\right)^{-1} \frac{1+2q/3}{1+2q/5}\langle L \rangle 
\end{equation}
and from (\ref{e16}),(\ref{e16a}) we have
\begin{equation}\label{eq5}
L^*=\left( \alpha+\frac{7}{2}\right)^{-1} \frac{1+2q/5}{1+2q/7}\frac{\langle L^2 \rangle}{\langle L \rangle}, 
\end{equation}
\begin{equation}\label{eq6}
L^*=\left( \alpha+\frac{9}{2}\right)^{-1} \frac{1+2q/7}{1+2q/9}\frac{\langle L^3 \rangle}{\langle L^2 \rangle}. 
\end{equation}
Equating these formulas we obtain two equations with two unknown quantities $q$ and $\alpha$. Combining them we obtain
an equation of fourth degree for $q$.

If some set of parameters $q_1$, $\alpha_1$ and $L^*_1$ provides a suitable fit of sample LF by function (\ref{e15})
one can calculate the values of $\langle L \rangle$, $\langle L^2 \rangle$ and $\langle L^3 \rangle$ for this function
and these parameters according to (\ref{eq1a}) and derive an equation for $q$. This equation of fourth degree must have 
at least one other solution with the exception of special case of multiple root. So, this solution provides other set 
of parameters $q_2$, $\alpha_2$ and $L^*_2$ with the same $\langle L \rangle$, $\langle L^2 \rangle$, $\langle L^3 \rangle$ 
and near the same values of higher moments of a frequency distribution. 

If $q_2>0$ then this set also provides a good fit of the
sample LF. Both distributions are practically coincident, despite
the fact that they are characterised by considerably different
sets of parameters. Naturally, if the distance between roots is small this leads to a complex form of
the 1$\sigma$ confidence region in the three-dimensional parameter
space and the strong correlation between the parameters for all methods of fitting, i.e. the maximum 
likelihood method (MLM) \citep{ref:H,ref:F}. The marginal errors
of these parameters obtained from the orthogonal projection
on ($L^*,\alpha$) plane are also large.

Therefore, the MLM and other methods of fitting are almost useless to find all three
parameters of the luminosity function (\ref{e15}) in the case of two positive roots of equation for $q$.
This situation occurs in processing the luminosity data for the sample of LCG used by \citep{PII}. 
However, if
the value of $q$ is fixed e.g. by the condition (\ref{e11}) the MLM reproduces quite well the
values of other two parameters. 

Note that as a by-product we obtain the method of estimating the values of parameters
$q$, $\alpha$ and $L^*$ that fit the sample LF by function (\ref{e15}). One can calculate the
values of $\langle L \rangle$, $\langle L^2 \rangle$ and $\langle L^3 \rangle$ for the sample
and use them in Eqs. (\ref{eq4}-\ref{eq6}).
                                          
\subsection{Luminosity function for the log-normal distribution of initial luminosities}\label{s:ln}
\begin{figure}[tb]
\includegraphics[width=\columnwidth]{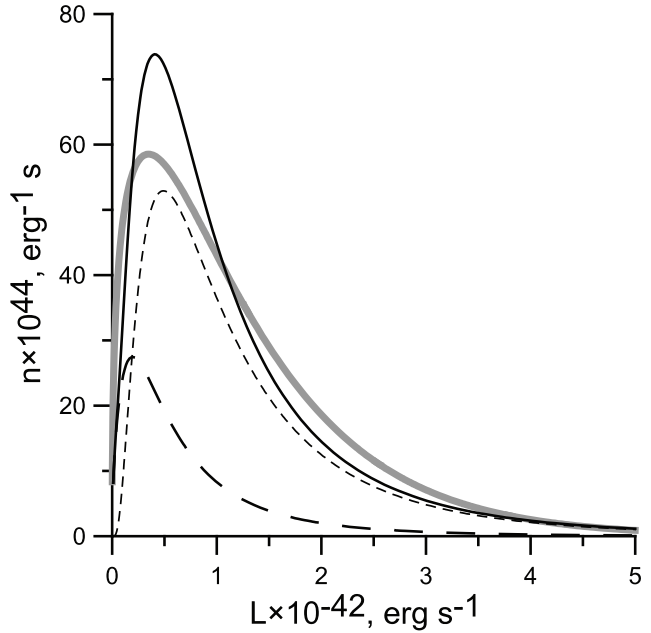}
\caption{The LF $n(L)$ (solid line) and the summands in (\ref{eq1}) 
(dashed lines with short and long dashes) for distribution (\ref{enn})
with parameters $q=0.48$, 
$\tilde{L}=4.9\cdot10^{41}$ erg s$^{-1}$ and 
$a=0.73$ are drawn by thin black
lines. The solid thick grey line corresponds to the distribution (\ref{e15}) with 
$q=0.48$, $L^*=8.5\cdot10^{41}$ erg s$^{-1}$, $\alpha=-0.88$
plotted in the Fig. \ref{f1}. Both distributions provides the best fit of LF of the sample of 795 LCGs by different
functions}
\label{f2}
\end{figure}

After substituting (\ref{en1}) into (\ref{eq1}) one can find the sample LF with log-normal initial LF
\begin{align}\label{enn}
n(L)=&\tilde{L}^{-1}\left(1+\frac{2q}{3}\right)^{-1}\times \nonumber\\
&\times\left[\left(\frac{a}{\pi}\right)^{1/2}\exp\left(-\frac{1}{4a}\right)\exp\left(-a\ln^2(L/\tilde{L})\right)+\right. \nonumber\\
&+\frac{q}{2}\exp\left(-\frac{3}{16a}\right)\left(\frac{L}{\tilde{L}}\right)^{1/2}\times \nonumber\\
&\left.\times\textrm{erfc}\left(\sqrt{a}\ln\left(\frac{L}{\tilde{L}}\right)+
\frac{1}{4\sqrt{a}}\right)\right].
\end{align}
Here
\begin{equation}\label{enn1}
\textrm{erfc}(x)=1-\textrm{erf}(x)=\frac{2}{\sqrt{\pi}}\int_x^{\infty}e^{-t^2}dt.
\end{equation}

In Fig. \ref{f2} this function and its summands are plotted for the set of the parameters 
$\tilde{L}=4.9\cdot10^{41}$ erg s$^{-1}$, $a=0.73$, providing the best fit of LF of the
sample of 795 LCGs. The best fit by the function (\ref{e15}) with (\ref{e11}) is shown by thick 
grey line. The first summand corresponds to the log-normal distribution (\ref{en1}).

The sample average values of $L^s$ are
\begin{equation}\label{enn2}
\langle L^s \rangle =\frac{1+\frac{2q}{3+2s}}{1+\frac{2q}{3}}\exp\left(\frac{s^2+2s}{4a}\right) \tilde{L}^s.
\end{equation}
From (\ref{enn2}) we get
\begin{equation}\label{enn3}
\tilde{L}=\frac{1+2q/3}{1+2q/5}\exp\left(-\frac{3}{4a}\right)\langle L \rangle,
\end{equation}
\begin{equation}\label{enn4}
\tilde{L}=\frac{1+2q/5}{1+2q/7}\exp\left(-\frac{5}{4a}\right)\frac{\langle L^2 \rangle}{\langle L \rangle},
\end{equation} 
\begin{equation}\label{enn5}
\tilde{L}=\frac{1+2q/7}{1+2q/9}\exp\left(-\frac{7}{4a}\right)\frac{\langle L^3 \rangle}{\langle L^2 \rangle}.
\end{equation} 
Dividing the product of (\ref{enn3}) and (\ref{enn5}) by the square of (\ref{enn4}) we obtain an 
equation of fourth degree for $q$ 
\begin{align}\label{enn6}
\left(1+\frac{2q}{7}\right)^3\left(1+\frac{2q}{3}\right)&=G\left(1+\frac{2q}{5}\right)^3\left(1+\frac{2q}{9}\right),\nonumber\\
G&=\frac{\langle L^2 \rangle^3}{\langle L \rangle^3 \langle L^3 \rangle}.
\end{align} 
It would seem that one can repeat the mention above argument. If
there exists a set of parameters $\tilde{L}_1$, $a_1$ and $q_1$ providing a good fit a LF of sample
by function (\ref{enn}) there have to exist other set of parameters $\tilde{L}_2$, $a_2$ and $q_2$ also
providing the same grade of fitting. Nevertheless, this situation never occurs for fitting by function
(\ref{enn}). For real luminosities the value of $G$ is near 1. For example for the sample of about 800
LCGs from \citet{PII} we have $G=1.0133$. In this case one root of (\ref{enn5}) is negative $q_2\approx -3$
and it does not interfere the fitting. Other root is positive at $G>1$ and negative at $G<1$. For the sample
of 795 LCGs we get $q_1=0.15$. This is less that the formula (\ref{e11}) predicts. But $q_1>0$ means that the function
(\ref{enn}) yields a better fit compared to the case of (\ref{enn}) at $q=0$ i.e. the log-normal distribution
(\ref{en1}).

There is no problem to use MLM or other methods of fitting to obtain a set of parameters of (\ref{enn}).

\section{Summary}\label{s:Sum}

The luminosity evolution affects the LF in the wavelength ranges and lines which are good indicators of the star 
formation, i.e. H$\alpha$ emission line, the UV continuum
and IR radiation because of the fading of galaxy luminosities on the time scale of several Myr 
after a starburst. 

We introduce and consider the initial luminosity 
function $\eta(L)$, the current luminosity function $\xi(L,T)$, the time-averaged 
luminosity function $\varphi(L)$ and the sample LF $n(L)$. For the typical time-dependence (\ref {e1}) we yield 
the relations (\ref {e2}) and (\ref {e6}) between them. We specify this relation for the luminosity evolution in 
the form (\ref {e9}) and analyze the obtained formula (\ref {e10}).

We consider the sample LF for the case when the initial LF is described by the gamma-distribution 
also known as the Schechter function (\ref {e12}). The sample LF is described by the 
distribution (\ref {e15}). There is a problem with this distribution, which complicates 
an approximation of the real data by the formula (\ref {e15}). The distribution 
(\ref {e15}) is quite similar for the different sets of parameters $q$, $\alpha$ and 
$L^*$ with the essentially different individual parameters. This fact obstructs the usage of the 
maximal likelihood method or other methods of approximation. The confidence 
region in the parameter space can be very prolate.
  
In the case of initial log-normal distribution (\ref {en1}) we obtain the sample LF in the form (\ref{enn}).
There is no problem to use MLM or other methods of fitting to obtain a set of parameters of this function.

Thus, we get two new types of LFs (\ref {e15}) and (\ref {enn}) to fit the LF of a sample of star-forming galaxies.
One can used it to fit the LF of a sample of star-forming galaxies.
These distributions have been used to fit the sample LF of LCGs studied by \citet{PII}. 
The results will be present in the next paper with an extended author list.

\end{document}